\documentclass[iopart,superscriptaddress,longbibliography,showpacs,preprint,11pt]{revtex4-1}

\usepackage{bm}
\usepackage{graphicx}
\usepackage{longtable}
\usepackage{amsmath}
\usepackage{amssymb}
\usepackage{color}

\usepackage{hyperref}

\begin{document}

\title{Controlling a three dimensional electron slab of graded Al$_{x}$Ga$_{1-x}$N}

\author{R.~Adhikari}
\email{rajdeep.adhikari@jku.at}
\affiliation{Institut f\"ur Halbleiter-und-Festk\"orperphysik, Johannes Kepler University, Altenbergerstr. 69, A-4040 Linz, Austria}

\author{Tian~Li}
\affiliation{Institute of Physics, Polish Academy of Sciences, al. Lotnik\'ow 32/46, PL-02 668 Warszawa, Poland}

\author{G.~Capuzzo}
\affiliation{Institut f\"ur Halbleiter-und-Festk\"orperphysik, Johannes Kepler University, Altenbergerstr. 69, A-4040 Linz, Austria}

\author{A. Bonanni}
\email{alberta.bonanni@jku.at}
\affiliation{Institut f\"ur Halbleiter-und-Festk\"orperphysik, Johannes Kepler University, Altenbergerstr. 69, A-4040 Linz, Austria}

\begin{abstract}
Polarization induced degenerate $n$-type doping with electron concentrations up to $\sim$10$^{20}$\,cm$^{-3}$ is achieved in graded Al$_{x}$Ga$_{1-x}$N layers ($x$: 0\%$\rightarrow$37\%) grown on unintentionally doped and on $n$-doped GaN:Si buffer/reservoir layers by metal organic vapor phase epitaxy. High resolution x-ray diffraction, transmission electron microscopy and electron energy loss spectroscopy confirm the gradient in the composition of the Al$_{x}$Ga$_{1-x}$N layers, while magnetotransport studies reveal the formation of a three dimensional electron slab, whose conductivity can be adjusted through the GaN(:Si) buffer/reservoir. 

\end{abstract}

\date{\today}


\maketitle

\section{Introduction}

The last two decades since the inception of the III-nitride based blue light emitting diodes (LEDs) \cite{Nakamura:1993_APL,Nakamura:1994_JVSTA}, have witnessed the emergence of III-nitrides as prominent material systems in state-of-the-art semiconductor technology \cite{Gutt:2012_APE, Morkoc:2008_book}. In particular, heterostructures based on Al$_{x}$Ga$_{1-x}$N/GaN represent the building-blocks for a number of current (opto)electronic devices, including blue and white LEDs \cite{Gutt:2012_APE, Morkoc:2008_book}, laser diodes and lasers \cite{Yoshida:2008_NatPhot}, high-power- \cite{Shur:1998_SSE,Yeluri:2015_APL} and high-electron-mobility-transistors (HEMTs)\,\cite{Mishra:2002_IEEE,Sun:2015_APL}.
Beside the tunability of the bandgap from the 3.4 eV of GaN to the 6.2 eV in AlN at room temperature, there is between GaN and AlN a 2.5\% lattice mismatch, allowing the pseudomorphic growth of Al$_{x}$Ga$_{1-x}$N layers with a biaxial tensile strain on GaN. This leads to a compressive strain along the $c$-axis of the wurtzite (wz) structure, which induces a macroscopic electric field. Additionally, due to the lack of inversion symmetry, wz-III-nitrides also have piezoelectric constants as large as $\sim$\,0.73\,Cm$^{-2}$-1.47\,Cm$^{-2}$ that exceed by over two orders of magnitude those of relevant zinc-blende semiconductors, whose values are in the range $\sim$\,0.01\,Cm$^{-2}$-0.1\,Cm$^{-2}$ [\onlinecite{Bernardini:1997_PRB,Bernardini:1997_PRL,Ambacher:2001_JPCM}]. 
With \textit{ab initio} calculations it was shown, that the macroscopic polarization in wurtzite nitrides is nonlinear\,\cite{Berdini:2001_PRB}. While the nonlinearity of the spontaneous polarization depends on the microscopic structure of the alloy, the nonlinear piezoelectric polarization is a bulk effect. The presence of this large spontaneous and induced polarization in Al$_{x}$Ga$_{1-x}$N/GaN heterostructures leads to the formation of a 2-dimensional electron gas (2DEG) with electron densities exceeding 10$^{13}$\,cm$^{-2}$\,[\onlinecite{Smorchkova:1999_JAP,Schmult:2006_PRB,Hsu:2001_JAP}], even in the absence of nominal doping. The polarization actually promotes the transfer of electrons from localized donor surface states at the barrier or even from the valence band to the quantum well at Al$_{x}$Ga$_{1-x}$N/GaN/Al$_{x}$Ga$_{1-x}$N interfaces. This property of III-nitride heterostructures is extensively exploited for the fabrication of high power--high frequency devices like HEMTs\,\cite{Palacios:2005_IEEE}. 
The polarization charge sheet across Al$_{x}$Ga$_{1-x}$N/GaN heterojunctions may be extended and spread into the Al$_{x}$Ga$_{1-x}$N by growing the latter with a compositional grading of Al. The spatially varying polarization $\textbf{P(r)}$ will then have a fixed charge volume with a density given by $\rho=\nabla\cdot\textbf{P(r)}$. Specifically, a compositional grading in the Al$_{x}$Ga$_{1-x}$N layer results in a non-vanishing divergence of the polarization, leading to the formation of a bulk three-dimensional (3D) polarization charge background, $i.e.$ to a polarization induced 3D electron slab (3DES)\,[\onlinecite{Jena:2002_APL,Jena:2003_PRB,Li:2013_APL}]. An analogous approach can also be exploited in a polarization doped hole system, where the grading may be designed to span from an Al rich to an Al depleted alloy\,\cite{Simon:2010_Science}. One of the challenges in developing nitride based deep-UV LEDs and lasers is the doping of the structures with donor or acceptor impurities like Si and Mg\,[\onlinecite{Walukiewicz:2001-PhysicaB}]. The Si activation energy in $n$-type Al(Ga)N can be as large as 280\,meV\,[\onlinecite{Borisov:2005_APL}], while for the $p$-type case, the Mg activation energy can be above 600\,meV for AlN\,[\onlinecite{Taniyasu:2006_Nature}]. Taking these facts into account, polarization induced doping of wide bandgap III-V nitride semiconductors may be viewed as an alternative and effective method to achieve high carrier concentration in otherwise insulating material systems, not only effective for the optimization of functional devices\,\cite{GolamSarwar:2015_APL,Li:2015_SST,Kivisaari:2013_APL,Park:2015_IEEE}, but opening new avenues for studies of collective phenomena such as integral and fractional quantum Hall effect\,\cite{Murzin:1998_PRL}, spin density waves, valley density wave states and Wigner crystallization in bulk systems\,\cite{Halperin:1987_JJAP,Kohmoto:1992_PRB,Koshino:2001_PRL,Koshino:2003_PRB,Brey:1989_PRB}.  
In previous works on polarization doped $n$-Al$_{x}$Ga$_{1-x}$N and on 3DES grown on a GaN buffer by means of molecular beam epitaxy (MBE)\,[\onlinecite{Jena:2002_APL,Jena:2003_PRB,Simon:2006_APL}] bulk carrier densities up to $\sim$ 10$^{18}$\, cm$^{-3}$ were reported. Li \textit{et al.} \cite{Li:2012_JAP} studied the reservoir effect of various substrates on the carrier density of a 3DES in graded Al$_{x}$Ga$_{1-x}$N (g-Al$_{x}$Ga$_{1-x}$N) layers grown by MBE. While Jena \textit{et al.} \cite{Jena:2003_PRB} reported electron mobilities as high as $\sim$ 10$^{3}$ \,cm$^2$/Vs in their 3DESs, the values given by Li \textit{et al.} [\onlinecite{Li:2012_JAP}] are $\sim\, 10^{2}$\,cm$^2$/Vs in 3DESs grown by MBE on semi-insulating and conducting GaN templates. 

Here, we report on 3DES in g-Al$_{x}$Ga$_{1-x}$N layers grown on differently conducting buffer layers by metal organic vapor phase epitaxy (MOVPE) on $c$-sapphire substrates and reaching carrier concentrations up to $\sim$ 10$^{20}$\, cm$^{-3}$. 
The samples are systematically characterized by atomic force microscopy (AFM), high resolution x-ray diffraction (HRXRD) and XRD reciprocal space mapping (RSM), transmission electron microscopy (TEM) also in high resolution (HRTEM) and scanning mode (STEM) in bright/dark field (BF/DF) and high angle annular dark field (HAADF), electron energy loss spectroscopy (EELS) and low temperature magnetotransport for Hall effect and magnetoresistance (MR). 
By combining the structural, chemical and magnetotransport studies we unravel the role of the buffer/reservoir on the carrier concentration in the 3DESs and also the effects of dislocation density, interface and alloy scattering on the mobility observed in the samples. 
We consider g-Al$_{x}$Ga$_{1-x}$N layers with 0.05~$\leq$~x~$\leq$~0.37 and deposited on GaN buffers (reservoir) having different electron concentration. In particular, the GaN-based reservoirs for the different samples are: a semi-insulating (SI) GaN:Mn buffer, an unintentionally conducting (u-)GaN buffer with background electron concentration $n_c\sim$ 10$^{16}$cm$^{-3}$ and a degenerate GaN:Si reservoir with $n_c\sim$ 10$^{19}$cm$^{-3}$.  As reference, we take an Al$_{0.25}$Ga$_{0.75}$N:Si/GaN and a g-Al$_{x}$Ga$_{1-x}$N layer grown on a SI GaN:Mn buffer. For all graded Al$_{x}$Ga$_{1-x}$N the gradient of Al concentration is $x$: 5\%$\rightarrow$30\% over the layer thickness.  Details on the S0-S5 studied samples structure and $n_c$ in the buffer/reservoir are provided in Table\,I. 

\section{Experimental}

The graded Al$_{1-x}$Ga$_{x}$N (g-Al$_{1-x}$Ga$_{x}$N) samples have been fabricated in an AIXTRON 200RF horizontal tube metal organic vapor phase epitaxy (MOVPE) reactor and deposited on a $c$-plane sapphire substrate using TMGa, TMAl, MnCp$_{2}$, SiH$_{4}$ and NH$_{3}$ as precursors for Ga, Al, Mn, Si and N respectively, with H$_{2}$ as carrier gas. Upon nitridation of the sapphire substrate, a low temperature GaN nucleation layer (NL) is grown at 540${^\circ}$C and then annealed at 1040${^\circ}$C. Then, a 1\,${\mu}$m thick GaN buffer is grown also at 1040$^{\circ}$C for sample S0-S2, while for samples S3-S5, a 1\,${\mu}$m thick semi-insulating GaN:Mn buffer with $\sim$\,0.6\% Mn cations concentration is deposited at 1040$^{\circ}$C. On the buffer, graded Al$_{1-x}$Ga$_{x}$N layers have been grown by continuously increasing the flow of the TMAl precursor from 0 to 35 standard cubic centimeters per minute (sccm) at a substrate temperature of 1000$^{\circ}$C. In samples S3 and S4, instead, a 500\,nm thick intermediate layer of degenerate \textit{n}-GaN:Si is inserted between the SI GaN:Mn buffer and the graded Al$_{1-x}$Ga$_{x}$N. All steps of the growth process are monitored by \textit{in situ} kinetic and spectroscopic ellipsometry.   

\begin{figure*}[ht]
	\centering
	\includegraphics[width=1.0\linewidth]{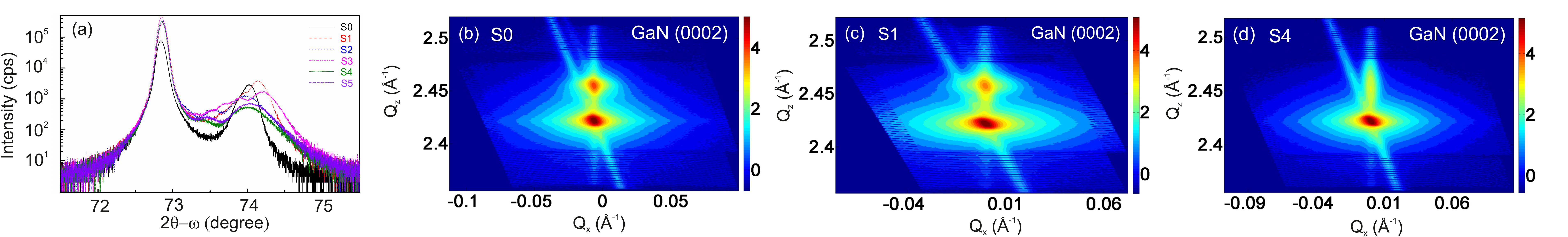}
	\caption{(Color online) (a): $2\theta\,-\omega$ scan of the (0004) reflection for samples S0-S5. The multiple peaks related to the Al$_{x}$Ga$_{1-x}$N (0004) Bragg reflection for samples S1-S5 point to a gradient in the Al content of the Al$_{x}$Ga$_{1-x}$N layers; (b)-(d) reciprocal space map of the symmetric (0002) reflection of GaN and Al$_{x}$Ga$_{1-x}$N for samples S0, S1, and S4, respectively.}
	\label{fig:fig1}
\end{figure*}

The HRXRD and RSM measurements have been carried out with a PANalytical's X'Pert PRO Materials Research Diffractometer (MRD) equipped with a hybrid monochromator with a 1/4$^{\circ}$ divergence slit. The diffracted beam is measured with a solid-state PixCel detector used as 256-channels detector with a 11.2\,mm anti-scatter slit.  
 In the radial $2\theta\,-\omega$ scans for $2\theta$ values between $30^\circ$ and $125^\circ$ (not shown here) only reflections from the $c$-Al$_{2}$O$_{3}$ substrate, from the GaN buffer/reservoir and from g-Al$_{x}$Ga$_{1-x}$N (Al$_{x}$Ga$_{1-x}$N:Si) are detectable. The radial HRXRD spectra of the symmetric (0004) reflection for the whole series of samples investigated in the present study are shown in Fig.\,\ref{fig:fig1}(a). While a sharp peak related to the Al$_{x}$Ga$_{1-x}$N (0004) reflection is detected for sample S0, for samples S1-S5 containing a g-Al$_{x}$Ga$_{1-x}$N the multiple peaks at the Al$_{x}$Ga$_{1-x}$N (0004) reflection are fingerprints of the gradient in the Al concentration. 

In order to gain a deeper insight into the distribution of Al in the samples, RSM of the symmetric (0002) reflections have been acquired and are reported in Figs.\,\ref{fig:fig1} (b)-(d) for samples S0, S1, and S4. The broadening in the direction Q$_x{}$ ($\omega$\,-\,scan) is determined by the tilt and finite lateral coherence length of mosaic blocks \cite{Moram:2009_RPP}.
In Figs.\,\ref{fig:fig1} (c)-(d), the elongation along the Q$_{z}$ direction points to a pseudomorphically grown Al$_{x}$Ga$_{1-x}$N layer with graded Al concentration on GaN. However, for sample S0 --as evidenced in Fig.\,\ref{fig:fig1} (b)-- this elongation of Al$_{x}$Ga$_{1-x}$N (0002)
along Q${_z}$ is not detected, confirming the presence of a non-graded Al$_{x}$Ga$_{1-x}$N:Si layer with a constant $x$$\sim$25\%.

\begin{figure*}[ht]
	\centering
	\includegraphics[width=1.0\linewidth]{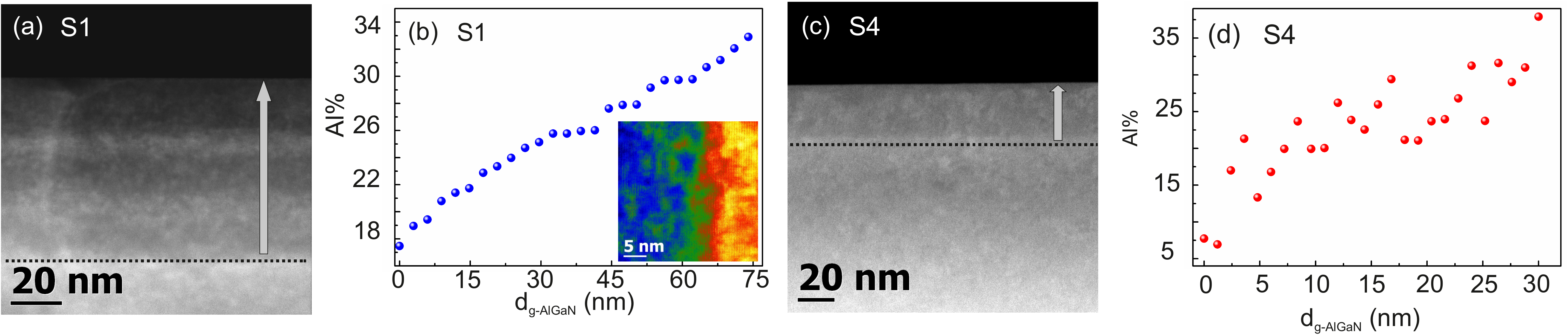}
	\caption{(Color online) Panels (a) and (c): TEM micrographs of S1 adn S4, respectively. Panels (b) and (d): EELS data for the graded layers in S1 and S4, respectively. Inset to panel (b): HAADF image confirming the presence of a $\approx$5\,nm wide intermixing region around the interface Al$_{x}$Ga$_{1-x}$N/GaN in S1.}
	\label{fig:fig2}
\end{figure*}

Measurements of TEM also in STEM mode are performed in a FEI Titan Cube 80-300 operating at 300\,keV and in a JEOL 2010F working at 200\,KeV.
Bright/dark-field BF/DF, HRTEM and HAADF are employed to analyse the structure of the sample, while chemical mapping is achieved through EELS analysis. 
Cross-section TEM specimens are prepared by mechanical polishing, dimpling and final ion milling in a Gatan Precision Ion Polishing System. 
The TEM micrograph of sample S1 reported in Fig.\,\ref{fig:fig2} (a) evidences the Al composition changes in the 75\,nm thick g-Al$_{x}$Ga$_{1-x}$N ($x$: 0\%$\rightarrow$37\%) layer, and the EELS data in Fig.\,\ref{fig:fig2} (b) confirm the  increase of the Al concentration in the g-Al$_{x}$Ga$_{1-x}$N corresponding to the nominal increment of Al flow rate during the process. On the other hand, for the 30\,nm thin g-Al$_{x}$Ga$_{1-x}$N ($x$: 0\%$\rightarrow$37\%) layer in S4, from the TEM image in Fig.\,\ref{fig:fig2} (c) it is not possible to appreciate the contrast between the g-Al$_{x}$Ga$_{1-x}$N regions with different Al content, but the monotonic increment of the Al concentration with the thickness is confirmed by the EELS data in Fig.\,\ref{fig:fig2} (d). This is in accord with the presence of a $\approx$5\,nm wide intermixing region around the interfaces between nominal variations in Al detected $via$ HAADF measurements as reported in the inset to Fig.\,\ref{fig:fig2} (b) for the interface Al$_{x}$Ga$_{1-x}$N/GaN.

For the Hall effect measurements and magnetoresistance studies, a four-probe van der Pauw geometry is employed. The Hall effect measurements are carried out in dc mode, while the MR studies are performed using a low frequency ac technique.  The ohmic contacts for all samples consist of a Ti/Au/Al/Ti/Au metal stack deposited in an e-beam evaporation chamber and then annealed at 750${^\circ}$C in N$_{2}$ atmosphere for 30 seconds.

\section{Results and discussion}

The above results of structural and chemical analysis of the samples confirm the presence of a controlled concentration gradient of Al in the wanted volume and the consequent non-vanishing divergence of the polarization is expected to determine the electrical properties of the samples.

The combination of spontaneous and piezoelectric polarization P(x) for an Al$_{x}$Ga$_{1-x}$N layer coherently strained to GaN is given by\,\cite{Ambacher:2001_JPCM}:
\begin{align}
P[x]=[2x+1.1875x^{2}+3.25x]\times 10^{13}
\end{align}
for a fixed concentration $x$ of Al. However, in the presence of a gradient in the content $x$ of Al over the Al$_{x}$Ga$_{1-x}$N layer thickness, the spatial change of the polarization leads to doping driven by dipole engineering and the induced charge N$_{Dp}$ is given by\,\cite{Wood:2008_Book}:
\begin{align}
N_{Dp}(x_{0},z_{0})=\frac{x_{0}}{z_{0}}(5.25+2.375x_{0}\dfrac{z}{x_{0}})
\end{align}
where x$_{0}$ is the concentration of Al at a position $z_{0}$ from the interface with the buffer and along the growth direction $z$ ($c$). 

By applying Eq.\,2 and by implementing the Al concentration as obtained from EELS measurements, the value of N$_{Dp}$ calculated at 30\,nm from the interface with the buffer for sample S4, is estimated to be (1.25$\times$10$^{19}$)\,cm$^{-3}$. A remarkable feature of this method of doping is that the doping level can be controlled by tuning the alloy composition and/or the thickness of the graded layer. The de Broglie wavelength $\lambda_{dB}=\frac{h}{\sqrt{2m^{*}k_{B}T}}$ is $\approx$17\,nm for non-degenerate carrier densities, while the Fermi-wavelength $\lambda_{F}$ for degenerate systems ($\sim$10$^{20}$\,cm$^{-3}$) is calculated to be $\sim$30\,nm. Any thickness of the graded layer which is smaller than these two characteristic lengths, leads to the formation of a quasi-2DEG. However, for wide slabs of graded regions with $z\geqslant\lambda_{dB},\lambda_{F}$, the electron gas is 3D.  

\begin{figure}[htbp*]
	\centering
	\includegraphics[width=0.5\linewidth]{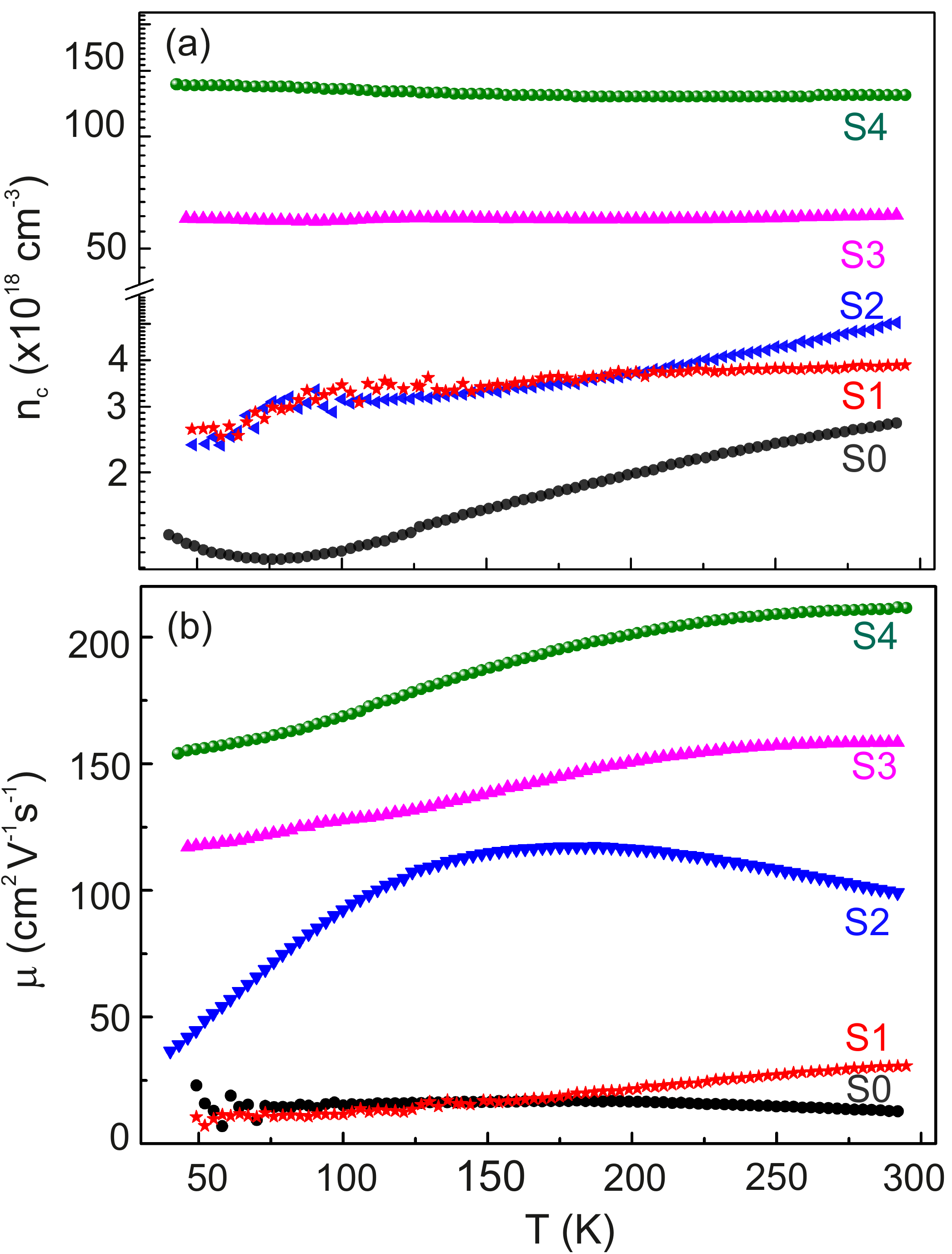}	
	\caption{(Color online) Hall carrier concentration (a) and mobilities (b) for samples S0-S4 over the temperature range 40\,K$\leq$T$\leq$300\,K.}
	\label{fig:fig3}
\end{figure}

In Fig.\,\ref{fig:fig3} (a) the electron concentration $n_c$ and in Fig.\,\ref{fig:fig3} (b) the Hall mobilities as a function of temperature in the range 40\,K$\leq$T$\leq$300\,K are reported for samples S0-S4 (S5 is semi-insulating). According to the curves in Fig.\,\ref{fig:fig3} (a), all the g-Al$_{x}$Ga$_{1-x}$N samples have carrier densities $\geq1\times10^{18}$\,cm$^{-3}$, i.e. well above the critical concentration for the metal-to-insulator transition (MIT) in \textit{n}-GaN:Si\,\cite{Stefanowicz:2014_PRB}.
By exploiting the Mott model, where the critical carrier concentration $N_{critical}$ is related to the Bohr radius of the relevant dopant a$_{B}$ as ${N_{critical}}^{1/3}a_{B}\approx\,0.26$ \cite{Dietl:2008_JSPJ}, we calculate $N_{critical}$ for \textit{n}-Al$_{0.20}Ga_{0.80}$N, \textit{n}-Al$_{0.30}$Ga$_{0.70}$N and \textit{n}-Al$_{0.40}$Ga$_{0.60}$N as (2$\times$10$^{18}$)\,cm$^{-3}$, (2.7$\times$10$^{18}$)\,cm$^{-3}$ and (4$\times$10$^{18}$)\,cm$^{-3}$, respectively. By comparing these values with the Hall data, we conclude that the reference Al$_{0.25}$Ga$_{0.75}$N:Si is doped degenerately and is above the MIT. The polarization doped g-Al$_{x}$Ga$_{1-x}$N samples also exhibit carrier concentrations far above the Mott limit for the corresponding Al$_{x}$Ga$_{1-x}$N systems doped with donor impurities like Si. 
 These results indicate that polarization induced doping in the studied graded Al$_{x}$Ga$_{1-x}$N drives such systems to a degenerate regime with the formation of a 3DES.

\begin{figure}[ht]
	\centering
	\includegraphics[width=0.5\linewidth]{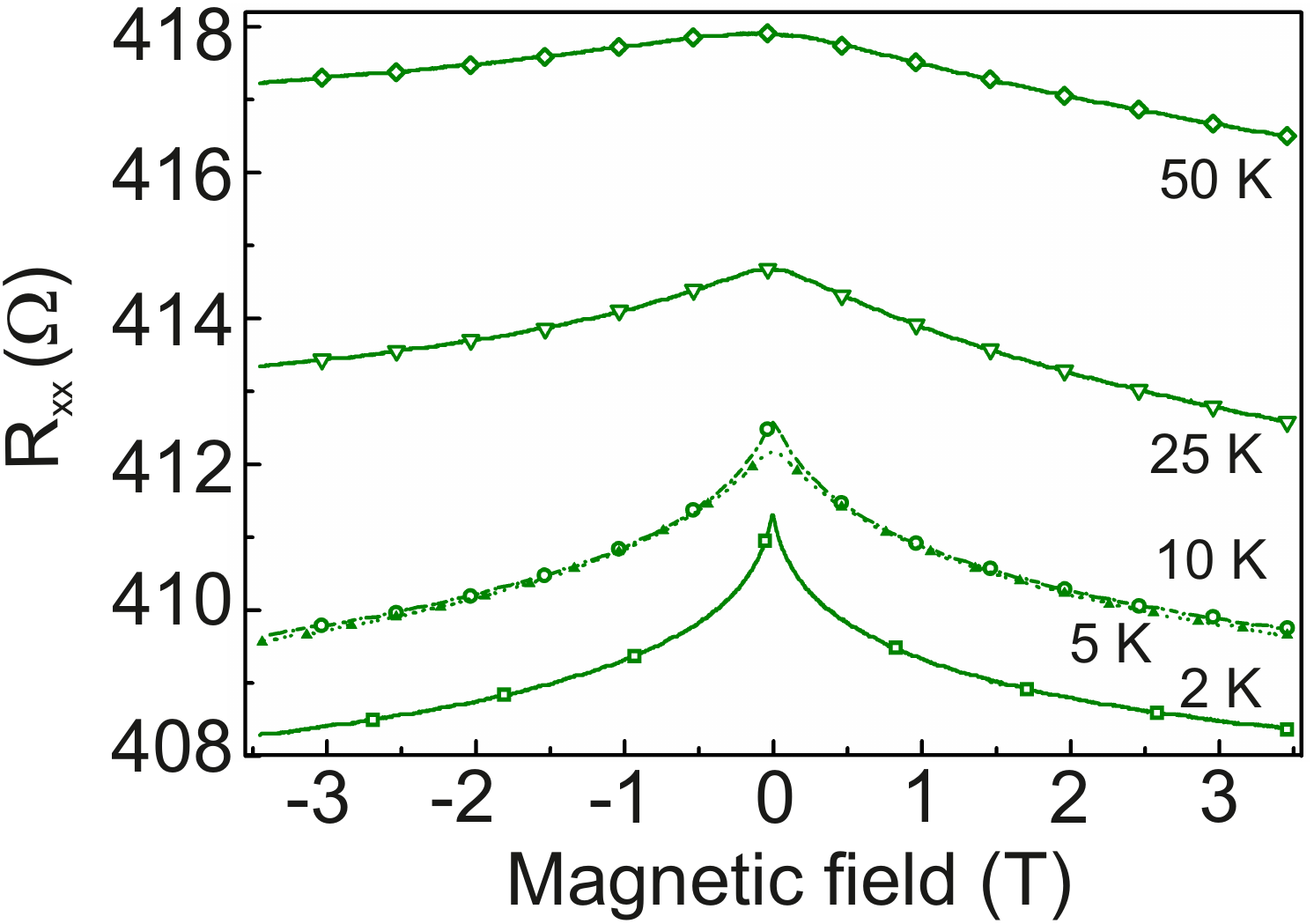}
	\caption{(Color online) Magnetoresistance of sample S4 for 2\,K$<$T$<$50\,K. The negative MR is a signature of quantum corrections to conductivity due to weak localization confirming the degeneracy of the 3DES.}
	\label{fig:fig4}
\end{figure}

Since all the samples discussed here have $n_c$ far above the Mott MIT limit, the graded Al$_{1-x}$Ga$_{x}$N layers are expected to be in a weak localization regime\,\cite{Stefanowicz:2014_PRB}. This is confirmed by the observation of negative MR (positive magnetoconductance) in the magnetoresistance data acquired as a function of the applied magnetic field and for various temperatures 2\,K$\leq$T$\leq$50\,K, as reported in Fig.\,\ref{fig:fig4} for sample S4.
If the 3DES is spread over a thickness $d_{0}$, then the sheet carrier density of the 3DES is calculated to be $n_{2D}=n_{3D}\times\,d_{0}$. For S4, this thickness is then 30\,nm, pointing to a 5\,nm depletion of the 3DES from the surface potential. 

The Hall parameters at 50\,K and 300\,K for samples S0-S5 are listed in Table\,I.  
 
 \begin{table*}[htbp]
 	\caption{Samples S0-S5: architecture, Hall parameters for the buffer/reservoir layer, carrier concentration ($n_{c}$), mobility ($\mu$) and room temperature resistivity ($\rho$) obtained from Hall effect measurements.}
 	\begin{tabular}{l l l l l l l l l}
 		\hline \hline
 		Sample  \hspace{0.1cm} & Structure \hspace{0.3cm} &  $n_c$ (reservoir)*   \hspace{0.2cm}  & $n_{c}$ (50~K) \hspace{0.3cm}  &  $n_{c}$ (300~K) \hspace{0.3cm}   &    $\mu$ (50~K) \hspace{0.3cm} & $\mu$ (300~K) \hspace{0.3cm} & $\rho$ (10$^{-2}$)     \\
 		\hspace{0.1cm} 	&  (nm)  \hspace{0.3cm} &  (cm$^{-3}$) \hspace{0.2cm} &  ($10^{18}$ cm$^{-3}$)  \hspace{0.3cm} & ($10^{18}$ cm$^{-3}$) \hspace{0.3cm} 	&   (cm$^{2}/$Vs) \hspace{0.3cm} & (cm$^{2}/$Vs) \hspace{0.3cm} & ($\Omega$$\cdot$cm) \\ \hline
 		S0  &	 Al$_{x}$Ga$_{1-x}$N:Si/                 &  $3\times$\,10$^{16}$   &  2.66   &  5.11  	&   15.88       &  13                	& 9.68    \\
				&  /u-GaN                    &                           & & & & & \\
				&  (100/1000)                    &                           & & & & & \\
		S1  &  g-Al$_{x}$Ga$_{1-x}$N/                  &  $3\times$\,10$^{16}$      &  3.16   	&  3.86  	&   15       		&  15.57     					&  11.03       \\
				&  /u-GaN                      &                           & & & & &             \\  
				&  (75/1000)                     &                           & & & & &             \\  
 		S2  &  g-Al$_{x}$Ga$_{1-x}$N/                  &  $3\times$\,10$^{16}$ &  1.27  		&  2.74  	&   44.5        &  97.74              &  2.3     \\
				&  /u-GaN                    &                           & & & & & 				\\
				& (35/1000)                    &                           & & & & & 				\\
 		S3  &  g-Al$_{x}$Ga$_{1-x}$N/ &     $1.2\times$\,10$^{19}$ & 60.2   	&  61.5 	&   117      		&  158.34             &   0.0635     \\  
				&  /GaN:Si/                 &          &                    & & & & & 			\\
				&  /SI-GaN:Mn                  &          &                    & & & & & 			\\
				& (75/500/1000)								&          &                    & & & & & 			\\
 		S4  &  g-Al$_{x}$Ga$_{1-x}$N/        &   $1.2\times$\,10$^{19}$    &  137   		&  128  	&   155      		&  215                &  0.027      \\
				& /GaN:Si/							&          &                    & & & & & 		\\
				&  /SI-GaN:Mn 							&          &                    & & & & & 		\\
				& (35/500/1000)						&          &                    & & & & & 		\\
 		S5  &  g-Al$_{x}$Ga$_{1-x}$N/               &  SI &  SI   		&  SI  		&   SI      		&  SI                 &   SI     \\ 
				& /SI-GaN:Mn   				 & & & & & &		\\ 
				& (75/1000)						& & & & & &		\\ \hline \hline 			
 	\end{tabular}
 	\label{tab:tab1}
 	\normalfont
	\begin{flushleft}		  	
		*Reservoir/buffer: layer on which g-Al$_{x}$Ga$_{1-x}$N and/or Al$_{x}$Ga$_{1-x}$N:Si are grown.	 	
	\end{flushleft} 
 \end{table*}

Sample S0 shows a carrier density of (2$\times$10$^{18}$)\,cm$^{-3}$ at 300\,K, which decreases with temperature. This sample is at the boundary of the MIT transition and for T$\leq$40\,K it can be driven to a non-degenerate electronic state. For samples S1 and S2, where the g-Al$_{x}$Ga$_{1-x}$N layer is grown on an u-GaN buffer with $n_c$$\sim$10$^{16}$cm$^{-3}$, the bulk carrier density at 300\,K is $\geq$(3$\times$10$^{18}$)\,cm$^{-3}$, and it remains substantially constant over the measured temperature range. However, for samples S3 and S4, grown on a degenerate GaN:Si reservoir with $n_{c}$=(1.2$\times$10$^{19}$)\,cm$^{-3}$, one can observe an enhanced carrier density, reaching (6$\times$10$^{19}$)\,cm$^{-3}$ for S3 and (1.4$\times$10$^{20}$)\,cm$^{-3}$ for S4. These values are greater than those observed for \textit{n}-GaN:Si\,[\onlinecite{Stefanowicz:2014_PRB}] and for 3DES in MBE g-Al$_{x}$Ga$_{1-x}$N/GaN systems \cite{Jena:2003_PRB,Li:2012_JAP}. The difference between S3 and S4 resides solely in the thickness of the g-Al$_{x}$Ga$_{1-x}$N layer. It is observed that sample S4 with a 35\,nm thickness of the graded layer shows higher carrier concentration when compared to S3, where the thickness of g-Al$_{x}$Ga$_{1-x}$N is 75\,nm. Also, from the observed carrier densities of the graded layers it can be concluded that the bulk concentration of carriers in the GaN reservior/buffer plays a  dominant role in determining the amount of charge carriers induced in a polarization doped g-Al$_{x}$Ga$_{1-x}$N system, where the graded layer grown on a degenerate GaN:Si shows a carrier concentration up to two orders of magnitude higher than when the g-Al$_{x}$Ga$_{1-x}$N layer is deposited on an u-GaN buffer. Furthermore, the absence of significant free charge carriers in the buffer, like $e.g.$ in S5, keeps g-Al$_{x}$Ga$_{1-x}$N semi-insulating. 

From Fig.\,\ref{fig:fig3} (b) we observe for samples S1-S4 carrier mobilities which are comparable\,\cite{Li:2012_JAP} or slightly lower than those observed in 3DES fabricated by MBE\,\cite{Jena:2002_APL,Jena:2003_PRB}. However,the Al concentration in the above mentioned works had a gradient in the range $x$:\,0\%$\rightarrow$20\%, while all the samples considered in this work have a Al gradient $x$:\,0\%$\rightarrow$37\%. A maximum mobility of 210 cm$^{2}$/Vs is achieved in S4, while for S1 a value of 15.57 cm$^{2}$/Vs at 300\,K is obtained. The samples S3 and S4 --with a rich reservoir of electrons in the GaN:Si buffer-- have higher mobilities compared to S1 and S2 deposited on u-GaN. From here it can be inferred, that the electronic properties (carrier concentration) of the reservoir/buffer directly affect the mobility in the 3DES. The dislocation density and the dislocation-related scattering of the charge carriers in the 3DES play also a role in determining the carrier mobility of the 3DES. From weak beam dark field image TEM data on sample S1, the screw- and edge-dislocation densities are estimated in (1.0$\times$10$^{8}$)\,cm$^{-2}$ and (8.1$\times$10$^{9}$)\,cm$^{-2}$, respectively,  while for sample S4, are of the order of (6.62$\times$10$^{8}$)\,cm$^{-2}$ and (1.26$\times$10$^{10}$)\,cm$^{-2}$. In the calculation, the average specimen thickness is established by counting the dark fringes in the BF image under two beam conditions, and the extinction distance of GaN used in the calculation is $\xi_{0}=$\,56\,nm\,[\onlinecite{Kong:2002_APL}].\\
Dislocations affect the carrier transport mostly through two mechanisms, namely $via$ the local deformation they induce in the lattice and the accumulation of static charges along the dislocation line or the core effect scattering. According to theoretical models, the main contribution to electron scattering due to dislocations in wz-III-nitride systems, originates from $a$-edge dislocations and from the edge component of ($a+c$) mixed dislocations. The contribution from $c$-screw dislocations was shown to have a negligible effect on the electron scattering (and consequently on the electron mobilities) in Al$_{x}$Ga$_{1-x}$N/GaN based structures\,\cite{Carosella:2008_JPCM}. From TEM measurements it is seen that the edge dislocation density in sample S4 is $\sim$1.55 times greater than in S1, but magnetotransport measurements reveal a Hall mobility of $\sim$\,210 cm$^{2}$/Vs in S4, much greater than the 15 cm$^{2}$/Vs of S1. This behaviour can be explained considering the carrier densities of the two samples. While in S4 we observe a carrier density of $\sim$(1.3$\times$10$^{20}$)\,cm$^{-3}$, the one measured in S1 is $\sim$(3.8$\times10^{18}$)\,cm$^{-3}$. With an increase in carrier density, the dislocation band located below the conduction band becomes narrower, reducing the strength of the scattering potential. Thus, despite a higher edge dislocation density, we observe a greater electron mobility in sample S4 compared to S1. Alloy disorder scattering and interface scattering additionaly affect the observed mobility in the 3DES samples. 

\section{Conclusions}

In summary, g-Al$_{x}$Ga$_{1-x}$N layers on buffer/reservoir GaN layers with different electron concentrations $n_c$ have been grown by MOVPE and thoroughly analysed. The compositional grading of the Al$_{x}$Ga$_{1-x}$N layer leads to the formation of a 3DES, whose Hall concentration depends on the electron concentration in the buffer/reservoir. The conductivity of the samples increases with increasing $n_c$ and this can be achieved either by augmenting the gradient range of the Al content while not varying the thickness of the graded layer, or by decreasing the layer thickness for a fixed range of the gradient, or by using a buffer/reservoir with a carrier concentration orders of magnitude higher than the one of u-GaN.  The observed magnetoresistance points to quantum corrections to the conductivity, due to weak localization\,\cite{Stefanowicz:2014_PRB}. With the present work it is shown that controllable g-Al$_{x}$Ga$_{1-x}$N leading to a 3DES can be fabricated in a MOVPE process. The realization of such degenerate Al$_{x}$Ga$_{1-x}$N-based systems opens wide perspectives for the fabrication of efficient electrodes in nitride-based deep UV LEDs and lasers, active layers for diodes, transistors and spin devices like nitride-based spin field effect transistors.

\section*{Acknowledgements}

This work was supported by the Austrian Science Foundation --
FWF (P22477 and P24471 and P26830), by the NATO Science for Peace Programme (Project No. 984735)
and by the EU $7^{\mathrm{th}}$  Framework Programmes:
CAPACITIES project REGPOT-CT-2013-316014 (EAgLE), and
FunDMS Advanced Grant of the European Research Council (ERC grant No.\,227690).


%

\end{document}